\documentclass[aps,pra,superscriptaddress,showpacs,tightenlines,cap,cs5size,winfonts,nospace,indent,fancyhdr,twocolumn]{revtex4-2}

\usepackage{amsmath}
\usepackage{changepage}
\usepackage{epsfig}
\usepackage{epstopdf}  
\usepackage[colorlinks,citecolor=blue,linkcolor=blue]{hyperref}
\usepackage{lipsum}
\usepackage{amssymb}
\usepackage{mathtools}
\usepackage{subfigure}
\usepackage{multirow}
\usepackage[table]{xcolor}
\usepackage{color}
\usepackage{graphicx}
\begin{document}

\title{Deterministic joint remote state preparation via a non-maximally entangled channel}
\author{Xuanxuan Xin}
\author{Shiwen He}
\author{Yongxing Li}
\author{Chong Li}
\thanks{lichong@dlut.edu.cn}
\affiliation{School of Physics, Dalian University of Technology, Dalian 116024,China}

\begin{abstract}
    Ideal deterministic quantum communication tasks require maximally entangled channels. The reality is that the maximally entangled channel is inevitably degraded to a generally entangled one because of various decoherence mechanisms, seriously deteriorating the performance of quantum communication. In most cases, entanglement purification and distillation are utilized to improve the entanglement and to construct the maximally entangled channel. In this paper, we proposed an alternative scheme to realize deterministic joint remote state preparation using a non-maximally entangled channel. Instead of spending additional entanglement resources in advance for entanglement purification or distillation, only two non-entangled ancillaries are employed in this scheme. Whether the employed quantum channel is a maximally entangled channel or a generally entangled one, remote state preparation would never fail theoretically in this investigation. This protocol provides a feasible way for the construction of practical quantum networks.
\end{abstract}

\maketitle

\section{Introduction}

Quantum communication takes advantage of the laws of quantum mechanics to transfer information encoded in quantum states between distant sites, the security of which is unreached by classical communication \cite{PhysRevLett.67.661, PhysRevLett.69.2881, Duan2001, Gisin2007}. Because of its unprecedented superiority over traditional programs, numerous quantum communication processes (QCP) have begun springing up in recent decades to deliver information more rapidly and safely \cite{PhysRevLett.70.1895, PhysRevA.58.4394, PhysRevA.58.4373, PhysRevLett.85.441, PhysRevLett.85.5635, PhysRevLett.95.010503, PhysRevLett.76.4656, PhysRevLett.93.210501, PhysRevLett.92.187901, PhysRevLett.89.187902, PhysRevA.69.052319, PhysRevA.71.044305}. Remote state preparation (RSP) \cite{PhysRevA.62.012313, PhysRevA.63.014302, PhysRevLett.87.077902} is one of these secure and efficient communication programs. In these schemes, a known quantum state coding messages is prepared in a remote place by the prior shared entanglement. The sender holds all prepared information in the standard RSP protocol \cite{PhysRevA.62.012313, PhysRevA.63.014302, PhysRevLett.87.077902}. Afterward, joint remote state preparation (JRSP) has been developed to improve the security further \cite{Xia_2007, Nguyen_2008,wang2013multiparty, AN20143582}. The difference between JRSP and RSP is the number of communicators and the location of the information to be prepared. In JRSP, the senders are changed from single to multiple. And the information is shared by more senders instead of only one sender. It restrains RSP to be accomplished only if all senders cooperate, thus further improving the security of communication.

In the original RSP protocol, the researcher successfully prepared a known real coefficient quantum state remotely by using the maximally entangled state \cite{PhysRevA.62.012313}. Unfortunately, when one tries to prepare a known complex coefficient quantum state remotely with the same channel, he found that the RSP scheme became probabilistic \cite{PhysRevA.63.014302}. To overcome this obstacle, many researchers have built various deterministic complex coefficient RSP (DRSP) protocols through maximally entangled channels \cite{PhysRevLett.90.057901, PhysRevA.73.022340, PhysRevA.81.042301, Nguyen_2011, Dakic2012, PhysRevA.98.042329}. However, it is lighthearted to prepare entanglement with the current experimental techniques whereas still challenging to produce the maximal entanglement for deterministic communication \cite{Lin2020, Lago-Rivera2021, Wein2022}. Moreover, the entanglements of the physical channels always deteriorate to the non-maximal entanglements due to inevitable decoherence \cite{PhysRevLett.78.3366, PhysRevLett.86.2913, PhysRevLett.93.230501} and ambient noises \cite{LINDNER2004321, PhysRevLett.120.084101, PhysRevLett.121.208301}, leading to the QCP degraded from deterministic to probabilistic, like probabilistic teleportation \cite{PhysRevA.62.024301, PhysRevA.61.034301, PhysRevA.68.022310, PhysRevA.91.012344}. Similarly, if the entangled channel is generalized from a maximally entangled channel to a non-maximally entangled channel, these DRSP protocols become probabilistic RSP (PRSP) schemes \cite{PhysRevLett.90.057901, PhysRevA.73.022340, PhysRevA.81.042301, Nguyen_2011, Dakic2012, PhysRevA.98.042329}. Usually, entanglement concentration and purification are utilized to convert the non-maximal entanglement to the maximal entanglement, and then RSP becomes deterministic QCP \cite{PhysRevLett.88.187903, PhysRevLett.90.207901, PhysRevA.77.062325, PhysRevA.85.012307, PhysRevA.100.052306, PhysRevLett.84.4002, Pan2003, Reichle2006, PhysRevLett.110.260503, Krastanov2019optimized, PhysRevLett.127.040502}. However, this strategy craves additional quantum resources, i.e., extra entanglement, to enhance the channel entanglement before communication, which is not only technically challenging but also less efficient in conversion \cite{Pan2001, PhysRevLett.90.067901, PhysRevA.72.012338, PhysRevApplied.10.054058}.

To this end, we offer an alternative solution here to achieve DJRSP in the degenerated entanglement environment. With the aid of ancillaries, a general quantum state is deterministically prepared in a remote place via a non-maximally entangled channel. We have addressed the communication failure resulting from the attenuated entanglement. Unit success probability is always achieved irrespective of the parameters of the preshared partial entanglement. There is no need to spend additional quantum resources in advance to improve the employed entanglement by purification and distillation, which significantly reduces the cost as well as the practicality of the operation.  Moreover, it is suitable for preparing quantum states of arbitrary dimensions, not only for preparing quantum states of 2, 4, and 8 dimensions \cite{geramita1979orthogonal, PhysRevA.65.022316}. Also, this deterministic communication is implemented in just one execution, without multiple repetitions \cite{PhysRevA.91.012344}. Compared to the non-standard protocol in 2018 \cite{PhysRevA.98.042329}, we have added an auxiliary particle and corresponding operations. But we solved the imperfect preparation when the channel is a non-ideal non-maximal entangled state, which provides a feasible way to practicalize QCP.

\begin{figure*}[htbp]
    \centering
    \includegraphics[width=0.7\paperwidth]{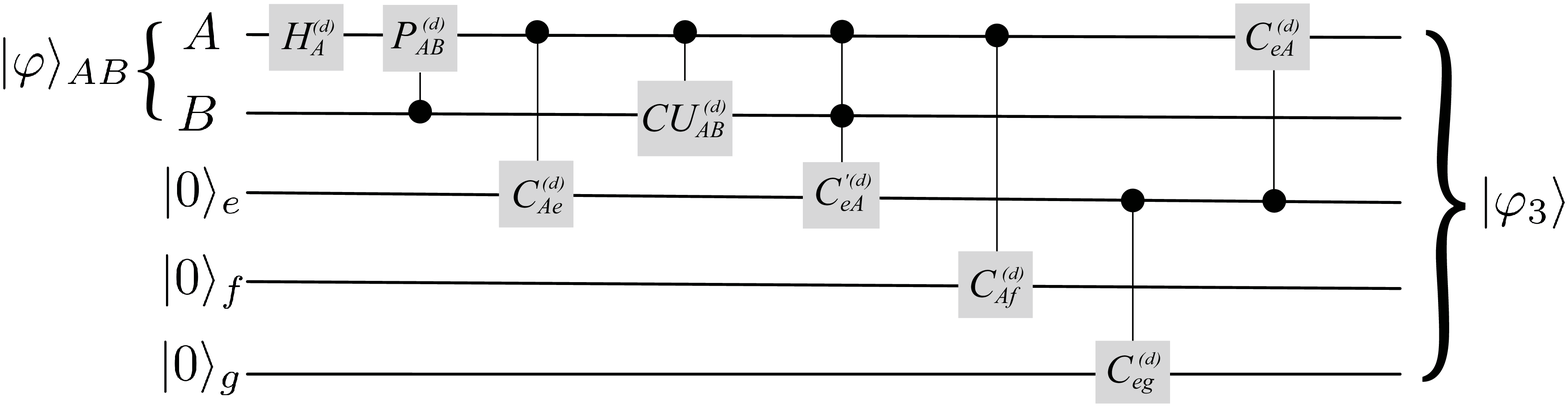}
    \caption{ The establishment of the non-maximum entanglement channel for DJRSP. At first, $Alice$ prepares the entangled state $|\varphi\rangle_{AB}=\sum_{k=0}^{d-1} a_{k} |kk\rangle_{AB}$ and the three auxiliary particles $\textbf{\emph{e}}$, $\textbf{\emph{f}}$, $\textbf{\emph{g}}$ set in the initial quantum state $|0\rangle_{i}$ severally. The solid line represents a qudit, the black circle represents the control qudit, and the gray rectangle represents the target qudit. $H_{ij}^{(d)}$ represents the Hadamard gate, $P_{ij}^{(d)}$ represents the phase gate, $C_{ij}^{(d)}$ represents the C-NOT gate, and $CU_{ij}^{(d)}$ represents the Controlled-U gate.}\label{Fig.1}
\end{figure*}

\section{DJRSP of a d-dimensional quantum state via a generally entangled state}

This proposed DJRSP protocol is suitable for the deterministic preparation of quantum states of arbitrary dimensions. A generalized quantum state is faithfully transferred from one place to the other with the aid of a non-maximally entangled channel and auxiliary particles. The JRSP does not fail regardless of whether the quantum channel used is an ideal maximally entangled channel or a non-maximally entangled one. The fresh execution of perfectly preparing a d-dimensional quantum state for a receiver by two senders is presented as follows. The communication parties involved in this DJRSP protocol are the two senders named $Alice$, $Charlie$, and one receiver named $Bob$. Suppose $Alice$ and $Charlie$ aims to prepare a d-dimensional quantum state $|\psi\rangle$ for Bob, where
\begin{eqnarray}    
     |\psi\rangle=\sum_{k=0}^{d-1} x_{k} |k\rangle=\sum_{k=0}^{d-1} |x_{k}|e^{i\theta_{k}} |k\rangle, \label{equation 22}
\end{eqnarray}
where $x_{0}, x_{1}, \cdots ,x_{d-1}$ are complex numbers and satisfy the orthonormalization $\sum_{j=0}^{d-1} |x_{j}|^{2}=1$. $Alice$ knows information $x_{0}, |x_{1}|, \cdots, |x_{j}|$, $Charlie$ has the phase information $\theta_{0}, \theta_{1}, \cdots, \theta_{j}$. Two senders do not know the full information to be prepared. Only two senders cooperate can complete information be transmitted to Bob. The preliminaries are the preparation of a two-qudit generally entangled state $|\varphi\rangle_{AB}=\sum_{k=0}^{d-1} a_{k} |kk\rangle_{AB}$ ($a_{0}, a_{1}, \cdots, a_{d-1}$ are complex numbers and satisfy the orthonormalization $\sum_{j=0}^{d-1} |a_{j}|^{2}=1$), two single-qudit quantum states $|0\rangle_{e}$, $|0\rangle_{g}$, and a single-qubit quantum state $|0\rangle_{f}$ by $Alice$. The initial quantum state of the whole system is described by:
\begin{eqnarray}
    \begin{aligned}  
    |\varphi_{0}\rangle_{ABefg}=\sum_{k=0}^{d-1} a_{k} |kk\rangle_{AB}|0\rangle_{e}|0\rangle_{f}|0\rangle_{g}.
    \end{aligned}
\end{eqnarray}

\textbf{Step (\uppercase\expandafter{\romannumeral1})} $Alice$ performs a generalized Hadamard gate $H^{(d)}_{i}=\frac{1}{\sqrt{d}}\sum_{r=0}^{d-1}e^{\frac{2\pi ikr}{d}}|r\rangle \langle k|$ on qudit $\textbf{\emph{A}}$ and a phase gate $P^{(d)}_{ij}=\frac{1}{\sqrt{d}}\sum_{k,r=0}^{d-1}e^{\frac{-2\pi ikr}{d}}|rk\rangle \langle rk |$ on qudits $\textbf{\emph{A}}$ and $\textbf{\emph{B}}$. Then she performs a generalized C-NOT gate $C^{(d)}_{ij}=(|0\rangle_{ii} \langle 0|) \otimes I_{j}  
+\sum_{k=1}^{d-1}\sum_{k^{'}=0}^{d-1}(|k\rangle_{ii} \langle k|) \otimes (|k^{'} \oplus r \rangle_{jj} \langle k^{'}|)$ on qudits $\textbf{\emph{A}}$ and $\textbf{\emph{e}}$. Here $\oplus$ denoting an addition modulo D. And it is specified that $|0 \oplus r \rangle_{j}$ = $|r \rangle_{j}$ when $k^{'}=0$.
\begin{eqnarray}
    \begin{aligned}  
    |\varphi_{1}\rangle_{ABefg}=&C^{(d)}_{Ae}P^{(d)}_{AB}H^{(d)}_{A}|\varphi_{0}\rangle_{ABefg}\\
   \sim&\sum_{r,k=0}^{d-1} a_{k} |rkr\rangle_{ABe}|0\rangle_{f}|0\rangle_{g}.
    \end{aligned} \label{equation 2}
\end{eqnarray}

\textbf{Step (\uppercase\expandafter{\romannumeral2})} $Alice$ applies a high-dimensional Controlled-U operation $CU^{(d)}_{ij}$ qudits $\textbf{\emph{A}}$ and $\textbf{\emph{B}}$, then the quantum state of the system $|\varphi_{1}\rangle$ is unitarily transformed into

\begin{eqnarray}
    \begin{aligned} 
            |\varphi_{2}\rangle_{ABefg}=&CU^{(d)}_{AB}|\varphi_{1}\rangle_{ABefg}\\
            \sim& \sum_{r=0}^{d-1}\sum_{k=1}^{d-1}[a_{0}(|r0r\rangle+|rkr\rangle)\\
    &+\sqrt{a_{k}^{2}-a_{0}^{2}}|(r \oplus s)kr\rangle]_{ABe}|0\rangle_{f}|0\rangle_{g},
    \end{aligned}  
  \end{eqnarray}
where
\begin{eqnarray}
    \begin{aligned} 
        CU^{(d)}_{AB}=&\sum_{k,r=0}^{d-1}a_{0}/a_{k}(|rk\rangle_{ABAB}\langle rk|\\
        &+|(r \oplus s)k\rangle_{ABAB}\langle (r \oplus s)k|)\\
        &+\sqrt{1-a^{2}_{0}/a^{2}_{k}}(|(r \oplus s)k\rangle_{ABAB}\langle rk|\\
        &-|rk\rangle_{ABAB}\langle (r \oplus s)k|).
    \end{aligned}  
  \end{eqnarray}

\textbf{Step (\uppercase\expandafter{\romannumeral3})} $Alice$ employs several generalized C-NOT gates on these qudits. Here assuming that the quantum state $|i \oplus s \oplus r\rangle=|i \oplus d\rangle=|i\rangle $. After that, the system has been evolved in
\begin{eqnarray}
    \begin{aligned} 
            |\varphi_{3}\rangle_{ABefg}=&C_{eA}^{(d)}C_{eg}^{(d)}C_{Af}^{(d)}\{C_{eA}^{'(d)}|\varphi_{2}\rangle_{ABefg}\\
            \sim&\sum_{r,k=1}^{d-1}[a_{0}(|r000\rangle+|00rr\rangle\\
        &+|rk00\rangle+|0krr\rangle)_{ABeg}|1\rangle_{f}\\
       &+\sqrt{a_{k}^{2}-a_{0}^{2}}(|0k00\rangle+|rkrr\rangle)_{ABeg}|0\rangle_{f}]. \label{equation 5}
    \end{aligned}  
\end{eqnarray}
Note here that the C-NOT gate $C_{ij}^{'(d)}$ is slightly different from $C_{ij}^{(d)}$: the operation $C_{ij}^{'(d)}$ acts to change the quantum state of particle $\textbf{\emph{j}}$ when the quantum state of particle $\textbf{\emph{i}}$ is $|0\rangle$; the quantum state of particle $\textbf{\emph{j}}$ does not change when the quantum state of $\textbf{\emph{i}}$ is not $|0\rangle$. This is exactly the opposite of $C_{ij}^{(d)}$. The normalized form of the above quantum state $|\varphi_{3}\rangle_{ABefg}$ is as follows:
\begin{eqnarray}
    \begin{aligned} 
        |\varphi_{3}\rangle_{ABefg}=&\sqrt{1/d(d-1)}\sum_{r,k=1}^{d-1}[a_{0}(|r000\rangle\\
        &+|00rr\rangle+|rk00\rangle+|0krr\rangle)_{ABeg}|1\rangle_{f}\\
       &+\sqrt{a_{k}^{2}-a_{0}^{2}}(|0k00\rangle+|rkrr\rangle)_{ABeg}|0\rangle_{f}]. \label{equation 5}
    \end{aligned}  
\end{eqnarray}

\begin{figure}
    \centering
    \includegraphics[width=0.325\paperwidth]{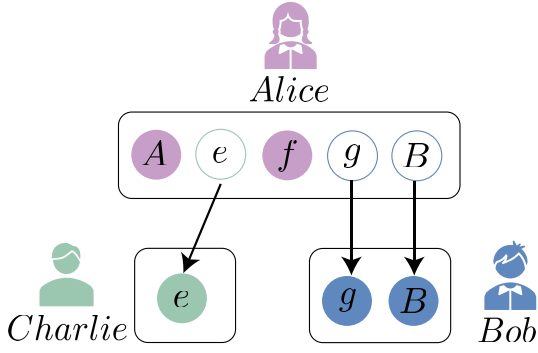}
    \caption{Particle distribution. $Alice$ has hold the five particles $\textbf{\emph{A}}$, $\textbf{\emph{e}}$, $\textbf{\emph{f}}$, $\textbf{\emph{g}}$ and $\textbf{\emph{B}}$ at this point. Next, $Alice$ distributes qubit $\textbf{\emph{e}}$ to $Charlie$, qubits $\textbf{\emph{g}}$ and $\textbf{\emph{B}}$ to $Bob$. qubits $\textbf{\emph{A}}$ and $\textbf{\emph{f}}$ remain in $Alice$'s position. Once the distribution is completed, an entanglement is created between three communicators.}\label{Fig.2}
\end{figure}

\textbf{Step (\uppercase\expandafter{\romannumeral4})}The quantum circuit diagram for above steps is illustrated in Fig. \ref{Fig.1}.  Next is the particle distribution: qudit $\textbf{\emph{A}}$ and qubit $\textbf{\emph{f}}$ are still placed in the location of $Alice$, qudit $\textbf{\emph{e}}$ is distributed to the second sender $Charlie$, qudits $\textbf{\emph{B}}$ and $\textbf{\emph{g}}$ are distributed to the receiver $Bob$. At this point, the entanglement has been established between the three communicators, as displayed in Fig. \ref{Fig.2}. $Alice$ performs a projective measurement on auxiliary qubit $\textbf{\emph{f}}$ under the orthogonal basis $\{|0\rangle,|1\rangle\}$. Then she tells this measurement outcome to $Charlie$ and $Bob$. There are two results of this measurement, $|0\rangle_{f}$ and $|1\rangle_{f}$.

If the measurement outcome is $|0\rangle_{f}$, the system is collapsed into 
\begin{eqnarray}
    \begin{aligned} 
        |\varphi_{4}\rangle_{ABeg}=\sum_{r,k=1}^{d-1}
       \sqrt{a_{k}^{2}-a_{0}^{2}}(|0k00\rangle+|rkrr\rangle)_{ABeg}.
    \end{aligned}  
\end{eqnarray}
$Bob$ applies the C-NOT gates $C_{gB}^{(d)}$ and $C_{Bg}^{(d)}$ on particles $\textbf{\emph{g}}$ and $\textbf{\emph{B}}$. Accordingly, the system has been involved into the following form
\begin{eqnarray}
    \begin{aligned} 
        |\varphi_{5}\rangle_{ABeg}=\sum_{r,k=1}^{d-1}
       \sqrt{a_{k}^{2}-a_{0}^{2}}(|000\rangle+|rkr\rangle)_{ABe}|0\rangle_{g}. \label{equation 8}
    \end{aligned}  
\end{eqnarray}
$Alice$ who knows $x_{0}$ and $|x_{1}|$ performs a projective measurement on qudit $\textbf{\emph{A}}$ based on the orthogonal basis $\{|\mu_{0}\rangle, |\mu_{1}\rangle, \cdots, |\mu_{p}\rangle, \cdots, |\mu_{d-1}\rangle\}$, where $|k\rangle=\frac{1}{\sqrt{d}}\sum_{p=0}^{d-1}|x_{k_{\textcolor{red}{p}}}||\mu_{p}\rangle$. Then $Alice$ informs the measurement outcome to $Charlie$ and $Bob$. If the outcome is $|\mu_{p}\rangle_{A}$, the quantum state of system is collapsed into
\begin{eqnarray}
    \begin{aligned} 
        |\varphi_{6}\rangle_{Beg}=&\sum_{r,k=1}^{d-1}\sqrt{a_{k}^{2}-a_{0}^{2}}(|x_{0_{p}}||00\rangle+|x_{r_{{p}}}||kr\rangle)_{Be}|0\rangle_{g}.
    \end{aligned}  
\end{eqnarray}
$Charlie$ measures qudit $\textbf{\emph{e}}$ based on the orthogonal basis $\{|\nu_{0}\rangle, |\nu_{1}\rangle, \cdots, |\nu_{d-1}\rangle \}$, where $|r\rangle=\frac{1}{\sqrt{d}}\sum_{q=0}^{d-1}e^{i\theta_{r_{q}}}|\nu_{q}\rangle$, then tells this menasurement result to $Bob$. If the result is $|\nu_{q}\rangle_{e}$, the quantum state of the system is collapsed into
\begin{eqnarray}
    \begin{aligned} 
        |\varphi_{7}\rangle_{Bg}=&\sum_{r,k=0}^{d-1}\sqrt{a_{k}^{2}-a_{0}^{2}}|x_{r_{p}}|e^{i \theta_{r_{q}}}|k\rangle_{B}|0\rangle_{g}.
    \end{aligned}  
\end{eqnarray}
$Bob$ performs the Hadamard gate $H^{(d)}_{B}$ and the phase gate $P^{(d)}_{B}=\frac{1}{\sqrt{d}}\sum_{r=0}^{d-1}e^{\frac{-2\pi ikr}{d}}|r\rangle \langle k|$ on qudit $\textbf{\emph{B}}$. Then he reconstructs the target state (Eq. (\ref{equation 22})) by adopting the corresponding unitary transformations on qudit $\textbf{\emph{B}}$ based on the measurements $|\mu_{p}\rangle_{A}$ and $|\nu_{q}\rangle$. It is found from the following formula that the target state is reconstructed perfectly by selecting the appropriate rotation operation on qudit $\textbf{\emph{B}}$.
\begin{eqnarray}
    \begin{aligned} 
        |\varphi_{8}\rangle_{Bg}=&\sum_{k=0}^{d-1}\sqrt{a_{k}^{2}-a_{0}^{2}}(\sum_{r=0}^{d-1}x_{r_{p}}|e^{i \theta_{r_{q}}}|k\rangle_{B})|0\rangle_{g}. \label{equation 11}
    \end{aligned}  
\end{eqnarray}

If the measurement outcome is $|1\rangle_{f}$, the system is collapsed into 
\begin{eqnarray}
    \begin{aligned} 
        |\varphi_{4}\rangle_{ABeg}=&\sum_{r,k=1}^{d-1}(|r000\rangle+|00rr\rangle\\
        &+|rk00\rangle+|0krr\rangle)_{ABeg}.
    \end{aligned}  
\end{eqnarray} 
$Bob$ applies the C-NOT gate $C_{Bg}^{(d)}$ on particles $\textbf{\emph{B}}$ and $\textbf{\emph{g}}$. Accordingly, the system has been involved into the following form
\begin{eqnarray}
    \begin{aligned} 
        |\varphi_{5}\rangle_{ABeg}=&\sum_{r,k=1}^{d-1}(|r000\rangle+|00rr\rangle\\
        &+|rk0r\rangle+|0kr0\rangle)_{ABeg}.
    \end{aligned}  
\end{eqnarray}
$Alice$ who knows $x_{0}$ and $|x_{1}|$ performs a projective measurement on qudit $\textbf{\emph{A}}$ based on the orthogonal basis $\{|\mu_{0}\rangle, |\mu_{1}\rangle, \cdots, |\mu_{p}\rangle, \cdots, |\mu_{d-1}\rangle\}$. Then $Alice$ informs the measurement outcome to $Charlie$ and $Bob$. If the outcome is $|\mu_{p}\rangle_{A}$, the quantum state of system is collapsed into
\begin{eqnarray}
    \begin{aligned} 
        |\varphi_{6}\rangle_{Beg}=&\sum_{r,k=1}^{d-1}[|x_{0_{p}}|(|0rr\rangle+|kr0\rangle)\\
        &+|x_{r_{p}}|(|000\rangle+|k0r\rangle)]_{Beg}.
    \end{aligned}  
\end{eqnarray}
$Charlie$ applies the phase gate $P(\theta)_{e}^{(d)}=\sum_{r=0}^{d-1}e^{2i(\theta_{r_{q}}-\theta_{0_{q}})}|r\rangle_{ee} \langle r|$ on qudit $\textbf{\emph{e}}$ and measures it based on the orthogonal basis $\{|\nu_{0}\rangle, |\nu_{1}\rangle, \cdots, |\nu_{d-1}\rangle \}$. Then this measurement result is told to $Bob$.
\begin{eqnarray}
    \begin{aligned} 
        |\varphi_{7}\rangle_{Bg}=&\frac{1}{d^{2}}\sum_{r,k=0}^{d-1}|x_{r_{p}}|e^{i \theta_{r_{q}}}|k\rangle_{B}|r\rangle_{g}. \label{equation 15}
    \end{aligned}  
\end{eqnarray}
$Bob$ measures the qudit $\textbf{\emph{g}}$ based on the basis $\{|0\rangle, |1\rangle, \cdots, |r\rangle, \cdots, |d-1\rangle\}$. From the above formula, it is concluded that $Bob$ can reconstruct the target quantum state $\sum_{k=0}^{d-1}|x_{k}|e^{i \theta_{k}}|k\rangle_{B}$ by performing the corresponding rotation operations on qudit $\textbf{\emph{B}}$ based on the three measurements $|\mu_{p}\rangle_{A}$, $|\nu_{q}\rangle_{e}$ and $|r\rangle_{g}$.

$\textbf{Summary}$: Derived from Eq. (\ref{equation 5}-\ref{equation 15}), the successful probability of JRSP is 1. The calculation of this value is related to measurements. The first measurement is for particle $\textbf{\emph{f}}$. There are two different measurement results, $|0\rangle_{f}$ and $|1\rangle_{f}$. From the Eq. (\ref{equation 5}), the incident probability of $|0\rangle_{f}$ is $\sum_{k=1}^{d-1}(a_{k}^{2}-a_{0}^{2})$, and the incident probability of $|1\rangle_{f}$ is $2a_{0}^{2}$. Next is the measurement of particle $\textbf{\emph{A}}$.There are $d$ kinds of measurement results, and the probability of each one occurring is same. Finally, there is the measurement for particle $\textbf{\emph{e}}$. There are $d$ different measurement results, and the probability of each one occurring is equal. No matter what combination of the above three measurement results, $Bob$ always successfully reconstructs the target quantum state (Eq. (\ref{equation 22})), derived from Eq. (\ref{equation 11}) and Eq. (\ref{equation 15}). Therefore, the sum of the success probabilities of JRSP is $\sum_{k=1}^{d-1}(a_{k}^{2}-a_{0}^{2})+2a_{0}^{2}=\sum_{k=0}^{d-1}a_{k}^{2}=1$. Whether $|a_{0}|$, $|a_{1}|$, $\cdots$, $|a_{d-2}|$ and $|a_{d-1}|$ are equal, the probability of success is always 1 theoretically.

\section{DJRSP of a 2-dimensional quantum state via a generally entangled state}

The solution provided in this paper is elaborated in this section with the preparation of two-dimensional quantum states as an example. Suppose the target state prepared for $Bob$ by $Alice$ and $Charlie$ is a two-dimensional quantum state
\begin{eqnarray} 
    \begin{aligned}    
     |\psi\rangle&=x_{0}|0\rangle+x_{1}|1\rangle\\
                    &=x_{0}|0\rangle+|x_{1}|e^{i\theta}|1\rangle,\label{equation 1}
    \end{aligned}
\end{eqnarray} 
where $x_{0}$ is a real number, $x_{1}$ is a complex number, $x_{0}$ and $x_{1}$ satisfy the orthonormalization $x_{0}^{2}+|x_{1}|^{2}=1$. $Alice$ knows information $x_{0}$ and $|x_{1}|$, $Charlie$ has the phase information $\theta$. Two senders do not know the full information to be prepared. Only two senders cooperate can complete messages be transmitted to Bob  deterministically. $Alice$ prepares a two-qubit generally entangled state $|\varphi\rangle_{AB}=(\alpha|00\rangle+\beta|11\rangle)_{AB} (\alpha$, $\beta$ are complex numbers, $|\alpha|^{2}+|\beta|^{2}=1$ and $|\alpha|<|\beta|)$ and three single-qubit quantum states $|0\rangle_{e}, |0\rangle_{f}, |0\rangle_{g}$ beforehand. The initial quantum state of the whole system is described by:
\begin{eqnarray}
    \begin{aligned}  
    |\varphi_{0}\rangle_{ABefg}=(\alpha|00\rangle+\beta|11\rangle)_{AB}|0\rangle_{e}|0\rangle_{f}|0\rangle_{g}.
    \end{aligned}
\end{eqnarray}

\textbf{Step (\uppercase\expandafter{\romannumeral1})} $Alice$ executes a Hadamard gate $H^{(2)}_{i}=1/\sqrt{2}[(|0\rangle+|1\rangle)_{ii}\langle 0|+ (|0\rangle-|1\rangle)_{ii}\langle 1|]$ \cite{PhysRevA.87.012307, PhysRevA.89.022340, PhysRevLett.125.180504} on qubit $\textbf{\emph{A}}$ and a phase gate $P^{(2)}_{ij}=|00\rangle_{ijij}\langle 00|+|10\rangle_{ijij}\langle 10|+|01\rangle_{ijij}\langle 01|-|11\rangle_{ijij}\langle 11|$ on qudits $\textbf{\emph{A}}$ and $\textbf{\emph{B}}$., then performs a C-NOT gate $C_{ij}^{(2)}$ \cite{PhysRevLett.94.030501,PhysRevLett.93.250502, PhysRevA.90.013808, PhysRevA.79.022301, Yamamoto2003} on qubits $\textbf{\emph{A}}$ and $\textbf{\emph{e}}$.
\begin{eqnarray}
    \begin{aligned} 
    |\varphi_{1}\rangle_{ABefg}=&C^{(2)}_{Ae}P^{(2)}_{AB}H^{(2)}_{A}|\varphi_{0}\rangle_{ABefg}\\
                       =&1/\sqrt{2}(\alpha|000\rangle+\alpha|101\rangle\\ &+\beta|010\rangle+\beta|111\rangle)_{ABe}
                      |0\rangle_{f}|0\rangle_{g}. 
    \end{aligned}  
  \end{eqnarray}
The expression of two-dimensional C-NOT gate is
\begin{eqnarray}
    \begin{aligned} 
         C_{ij}^{(2)}=&(|0\rangle_{ii}\langle 0|) \otimes I_{j}\\
         &+(|1\rangle_{ii}\langle 1|) \otimes (|1\rangle_{jj}\langle 0|+|0\rangle_{jj}\langle 1|),  
        \end{aligned}  
    \end{eqnarray}       
where $\textbf{\emph{i}}$ is controll qubit, $\textbf{\emph{j}}$ is target qubit and \emph{I} is the unit operation. When the quantum state of controll qubit $\textbf{\emph{i}}$ is $|0\rangle$, the quantum state of target qubit $\textbf{\emph{j}}$ is unchanged; when the quantum state of controll qubit $\textbf{\emph{i}}$ is $|1\rangle$, Pauli operator $\sigma_{x}$ is performed on target qubit $\textbf{\emph{j}}$.

\textbf{Step (\uppercase\expandafter{\romannumeral2})} $Alice$ applies a Controlled-U gate $CU_{ij}^{(2)}$ \cite{PhysRevA.62.024301, PhysRevA.61.034301, PhysRevA.68.022310, PhysRevA.91.012344} on qubits $\textbf{\emph{A}}$ and $\textbf{\emph{B}}$, then the quantum state of the system $|\varphi_{1}\rangle$ is unitarily transformed into
\begin{eqnarray}
    \begin{aligned} 
    |\varphi_{2}\rangle_{ABefg}=&CU_{AB}^{(2)}|\varphi_{1}\rangle_{ABefg}\\
                       =&\alpha/\sqrt{2}(|000\rangle+|101\rangle\\
                       &+|010\rangle+|111\rangle)_{ABe}|0\rangle_{f}|0\rangle_{g}\\
                       &+\sqrt{(\beta^2-\alpha^2)/2}(|110\rangle+|011\rangle)_{ABe}|0\rangle_{f}|0\rangle_{g},
    \end{aligned}  
\end{eqnarray}
where 
\begin{eqnarray}
    \begin{aligned} 
        CU_{AB}^{(2)}=&|00\rangle_{ABAB}\langle 00|+|10\rangle_{ABAB}\langle 10|\\
                       &+\alpha/\beta(|01\rangle_{ABAB}\langle 01|+|11\rangle_{ABAB}\langle 11|)\\
                       &+\sqrt{1-\alpha^{2}/\beta^{2}}(|11\rangle_{ABAB}\langle 01|-|01\rangle_{ABAB}\langle 11|).
    \end{aligned}  
\end{eqnarray}

\textbf{Step (\uppercase\expandafter{\romannumeral3})} $Alice$ performs several C-NOT gates on the system.
\begin{eqnarray}
    \begin{aligned} 
        |\varphi_{3}\rangle_{ABefg}=&C_{eA}^{(2)}C_{eg}^{(2)}C_{Af}^{(2)}C_{eA}^{'(2)}|\varphi_{2}\rangle_{ABefg}\\
        =&\alpha/\sqrt{2}(|1000\rangle+|0011\rangle\\
        &+|1100\rangle+|0111\rangle)_{ABeg}|1\rangle_{f}\\
        &+\sqrt{(\beta^2-\alpha^2)/2}(|0100\rangle+|1111\rangle)_{ABeg}|0\rangle_{f}.
    \end{aligned}  
\end{eqnarray}

\begin{figure*}[htbp]
    \centering
    \includegraphics[width=0.8\paperwidth]{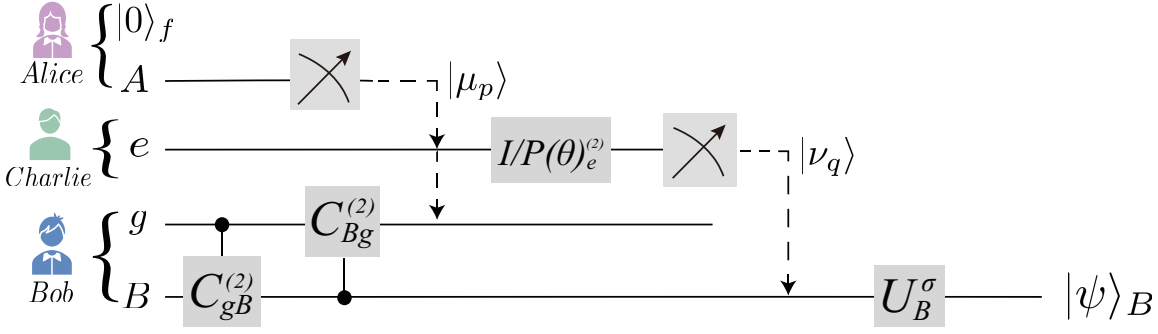}
    \vspace{-16em}
    \caption{Quantum circuit diagram of posterior measurements corresponding to the measurement result $|0\rangle_{f}$. A dashed line represents a classical dit. The solid arrow represent the projection measurement. The operation $U_{B}^{\sigma}$ represents the operation related to Pauli operator.}\label{Fig.3}
\end{figure*}
\textbf{Step (\uppercase\expandafter{\romannumeral4})} Next is the distribution of qubits to establish the entanglement between $Alice$, $Charlie$ and $Bob$. Qubits $\textbf{\emph{A}}$ and $\textbf{\emph{f}}$ are still placed in the location of $Alice$. Qubit $\textbf{\emph{e}}$ is distributed to the second sender $Charlie$. Qubits $\textbf{\emph{B}}$ and $\textbf{\emph{g}}$ are distributed to the receiver $Bob$.

\textbf{Step (\uppercase\expandafter{\romannumeral5})} To this point, three communicators have established the entanglement relationship. Next is to reconstruct the target state for $Bob$. Firstly, $Alice$ performs a projective measurement on auxiliary qubit $\textbf{\emph{f}}$ under the orthogonal basis $\{|0\rangle,|1\rangle\}$. There are two measurements, $|0\rangle_{f}$ and $|1\rangle_{f}$. Secondly, $Alice$ knowing the prepared information $x_{0}$ and $|x_{1}|$ performs a projective measurement on qubit $\textbf{\emph{A}}$ based on the orthogonal basis $\{|u_{0}\rangle=x_{0}|0\rangle+|x_{1}||1\rangle, |u_{1}\rangle=|x_{1}||0\rangle-x_{0}|1\rangle\}$. There are also two measurements, $|u_{0}\rangle_{A}$ and $|u_{1}\rangle_{A}$. Then she sends the above measurement results to $Charlie$ and $Bob$ respectively. $Charlie$ who knows the phase information $\theta$ performs a projective measurement on qubit $\textbf{\emph{e}}$ under the orthogonal basis $\{|\nu_{0}\rangle=|0\rangle+e^{i\theta}|1\rangle, |\nu_{1}\rangle=e^{-i\theta}|0\rangle-|1\rangle\}$. Then he informs $Bob$ of the measurement result. After $Bob$ has received the measurements from $Alice$ and $Charlie$, the target state $|\psi\rangle$ is reconstructed by performing the corresponding correction operations for the different measurements. Regardless of which of the eight measurements results \{$|0\rangle_{f}|u_{0}\rangle_{A}|\nu_{0}\rangle_{e}$, $|0\rangle_{f}|u_{0}\rangle_{A}|\nu_{1}\rangle_{e}$, $|0\rangle_{f}|u_{1}\rangle_{A}|\nu_{0}\rangle_{e}$, $|0\rangle_{f}|u_{1}\rangle_{A}|\nu_{1}\rangle_{e}$, $|1\rangle_{f}|u_{0}\rangle_{A}|\nu_{0}\rangle_{e}$, $|1\rangle_{f}|u_{0}\rangle_{A}|\nu_{1}\rangle_{e}$, $|1\rangle_{f}|u_{1}\rangle_{A}|\nu_{0}\rangle_{e}$, $|1\rangle_{f}|u_{1}\rangle_{A}|\nu_{1}\rangle_{e}$\}, the receiver can successfully reconstruct the target state with a 100\% probability. Refer to Appendix \ref{app2} for specific details.

\section{Conclusion}

While there has been a tremendous advancement in quantum communication so far, degradation of entanglement due to the  noisy environment is a sticky challenge to address in transporting messages perfectly. We reinvestigated probabilistic quantum communication protocols and devised a deterministic state preparation protocol. Any given d-dimensional quantum state is always prepared in a remote position with a 100\% success probability using the non-maximally entangled channel low requirements on preparation. Most previous resource-intensive schemes rely on entanglement purification to enhance the channel entanglement beforehand. But the partially entangled channel without any adjustment is employed to state preparation in this scheme. No additional quantum resources have to be consumed in advance to improve the intensity of the entanglement used. There is also no increase in the consumption of classical-information cost compared to previous programs. This well-designed protocol applies to transporting high-dimensional quantum states perfectly, which is constructive for conveying more information in integrated quantum networks. Transmitting quantum information between various high-dimensional quantum systems has significant advantages in robustness against errors and reduces quantum circuits complexities. This work overcomes the negative effects of diluted entanglement and significantly reduces the challenges of applying quantum technologies in practice. Future research should consider its application of other quantum communication schemes, such as quantum teleportation, quantum key distribution, quantum dense coding, quantum secure direct communication, etc.

\appendix

\section{The specific mathematical derivation following the measurement on qubit $\textbf{\emph{f}}$} \label{app2}

\subsection{The measurement outcome of $|0\rangle_{f}$}

If the measurement outcome is $|0\rangle_{f}$, the quantum state of the system is collapsed into
\begin{eqnarray}
    \begin{aligned} 
        |\varphi_{4}\rangle_{ABeg}=(|0100\rangle+|1111\rangle)_{ABeg}.
    \end{aligned} 
\end{eqnarray}
After being informed the measurement result $|0\rangle_{f}$ by $Alice$, $Bob$ apllies the C-NOT gate $C_{gB}^{(2)}$ and $C_{Bg}^{(2)}$ on particles $\textbf{\emph{g}}$ and $\textbf{\emph{B}}$. The system thus evolves into the following quantum state
\begin{eqnarray}
    \begin{aligned} 
        |\varphi_{5}\rangle_{ABeg}=(|010\rangle+|101\rangle)_{ABe}|1\rangle_{g}.
    \end{aligned} 
\end{eqnarray}

\textbf{Step (\uppercase\expandafter{\romannumeral1})} $Alice$ who knows $x_{0}$ and $|x_{1}|$ performs a projective measurement on qubit $\textbf{\emph{A}}$ based on the orthogonal basis $\{|u_{0}\rangle=x_{0}|0\rangle+|x_{1}||1\rangle, |u_{1}\rangle=|x_{1}||0\rangle-x_{0}|1\rangle\}$. Then $Alice$ sends the measurement result to $Charlie$ and $Bob$. Under this orthogonal basis, the quantum state $|\varphi_{5}\rangle$ can be written as follows,
\begin{eqnarray}
    \begin{aligned}  
    |\varphi_{5}\rangle_{ABeg}=&[|u_{0}\rangle_{A}(x_{0}|10\rangle-|x_{1}||01\rangle)_{Be}\\
    &+|u_{1}\rangle_{A}(|x_{1}||10\rangle+x_{0}|01\rangle)_{Be}]|1\rangle_{g}.
    \end{aligned} 
\end{eqnarray}
From the above equation, there are two kinds of measurement outcomes of qubit $\textbf{\emph{A}}$, $|u_{0}\rangle_{A}$ and $|u_{1}\rangle_{A}$. The next portion will be discussed separately based on these two different measurements.

{\subsubsection{\bfseries{The measurement outcome of $|u_{0}\rangle_{A}$}}}

If the outcome of the measurement is $|u_{0}\rangle_{A}$, the quantum state $|\varphi_{5}\rangle$ is collapsed into
\begin{eqnarray}
    |\varphi_{6}\rangle_{Beg}=(x_{0}|10\rangle-|x_{1}||01\rangle)_{Be}|1\rangle_{g}.
\end{eqnarray}

\textbf{Step (\uppercase\expandafter{\romannumeral2})} $Charlie$ who knows the phase information performs the phase gate $P(\theta)^{(2)}_{e}=(|0\rangle\langle 0|+e^{2i\theta}|1\rangle\langle 1|)_{e}$ on qubit $\textbf{\emph{e}}$ and operates a projective measurement on qubit $\textbf{\emph{e}}$ under the orthogonal basis $\{|\nu_{0}\rangle=|0\rangle+e^{i\theta}|1\rangle, |\nu_{1}\rangle=e^{-i\theta}|0\rangle-|1\rangle\}$.
\begin{eqnarray}
    \begin{aligned}  
    |\varphi_{6}\rangle_{Beg}=&[|\nu_{0}\rangle_{e}(x_{0}|1\rangle-x_{1}|0\rangle)_{B}\\
                      &+e^{i\theta}|\nu_{1}\rangle_{e}(x_{0}|1\rangle+x_{1}|0\rangle)_{B}]|1\rangle_{g}.
    \end{aligned} 
\end{eqnarray}
From the above equation, $Bob$ performs the Pauli operator $\sigma_{z}$ to get the target state $|\psi\rangle=x_{0}|0\rangle+x_{1}|1\rangle$ if the measurement outcome is $|\nu_{0}\rangle_{e}$; if the measurement outcome is $|\nu_{1}\rangle_{e}$, Bob performs the Pauli operator $\sigma_{x}$ correspondingly.

{\centering\subsubsection{\bfseries{The measurement outcome of $|u_{1}\rangle_{A}$}}}

If the measurement outcome is $|u_{1}\rangle_{A}$, the quantum state of the system is collapsed into
\begin{eqnarray}
    |\varphi_{6}\rangle_{Beg}=(|x_{1}||10\rangle+x_{0}|01\rangle)_{Be}|1\rangle_{g}.
\end{eqnarray}

\textbf{Step (\uppercase\expandafter{\romannumeral2})} $Charlie$ measures qubit $\textbf{\emph{e}}$ based on the orthogonal basis $\{|\nu_{0}\rangle, |\nu_{1}\rangle\}$.
\begin{eqnarray}
    \begin{aligned}  
        |\varphi_{6}\rangle_{Beg}=&|1\rangle_{g}[e^{-i\theta}|\nu_{0}\rangle_{e}(x_{0}|0\rangle+x_{1}|1\rangle)_{B}\\
                                                &+|\nu_{1}\rangle_{e}(-x_{0}|0\rangle+x_{1}|1\rangle)_{B}].
    \end{aligned} 
\end{eqnarray}
$Charlie$ sends the measurement result of qubit $\textbf{\emph{e}}$ to $Bob$ .If the measurement outcome is $|\nu_{0}\rangle$, Bob gets the target state $|\psi\rangle=x_{0}|0\rangle+x_{1}|1\rangle$; if the measurement outcome is $|\nu_{1}\rangle$, Bob performs the Pauli operator $\sigma_{z}$  accordingly to obtain the target state $|\psi\rangle$. The above operation of the measurement outcome $|0\rangle_{f}$ is demonstrated in Fig. \ref{Fig.3}.

\begin{figure*}[htbp]
    \centering
    \includegraphics[width=0.7\paperwidth]{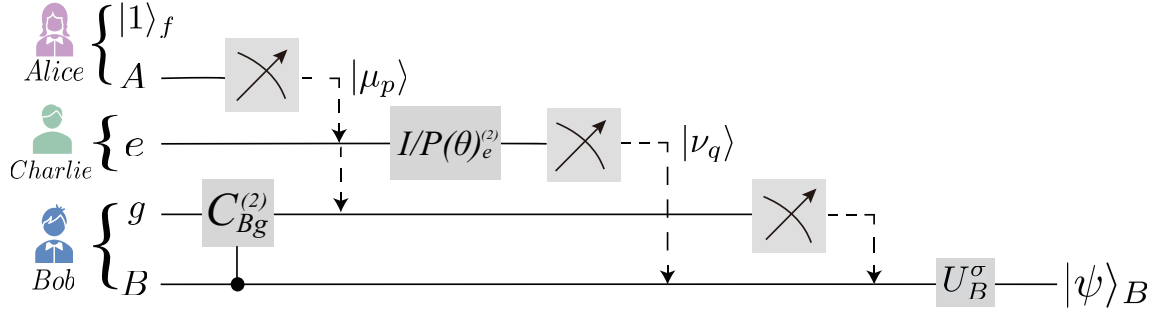}
    \caption{Quantum circuit diagram of posterior measurements corresponding to the measurement result $|1\rangle_{f}$.}\label{Fig.4}
\end{figure*}

\subsection{The measurement outcome of $|1\rangle_{f}$}

If the measurement outcome is $|1\rangle_{f}$, the quantum state of the whole system is collapsed into
\begin{eqnarray}
    \begin{aligned} 
        |\varphi_{4}\rangle_{ABeg}=&(|1000\rangle+|0011\rangle\\
        &+|1100\rangle-|0111\rangle)_{ABeg}.
    \end{aligned} 
\end{eqnarray}
After being informed the measurement result $|1\rangle_{f}$ by $Alice$, $Bob$ applies C-NOT gate $C_{Bg}^{(2)}$ on qubits $\textbf{\emph{B}}$ and $\textbf{\emph{g}}$. Accordingly, the system evolves into
\begin{eqnarray}
    \begin{aligned} 
        |\varphi_{5}\rangle_{ABeg}=&(|1000\rangle+|0011\rangle\\
        &+|1101\rangle-|0110\rangle)_{ABeg}.
    \end{aligned} 
\end{eqnarray}

\textbf{Step (\uppercase\expandafter{\romannumeral1})} $Alice$ who knows $x_{0}$ and $|x_{1}|$ performs a projective measurement on qubit $\textbf{\emph{A}}$ based on the orthogonal basis $\{|u_{0}\rangle, |u_{1}\rangle\}$. Under this orthogonal basis, the quantum state $|\varphi_{5}\rangle$ can be written as follows,
\begin{eqnarray}
    \begin{aligned} 
        |\varphi_{5}\rangle_{ABeg}=&|u_{0}\rangle_{A}(|x_{0}||011\rangle-|x_{0}||110\rangle
        +|x_{1}||000\rangle\\
        &+|x_{1}||101\rangle)_{Beg}
        +|u_{1}\rangle_{A}(|x_{1}||011\rangle\\
        &-|x_{1}||110\rangle
        -|x_{0}||000\rangle-|x_{0}||100\rangle)_{Beg}.
    \end{aligned} 
\end{eqnarray}
After the measurement, $Alice$ sends the measurement result to $Charlie$ and $Bob$. In the following, the various measurement results are discussed separately.

{\centering\subsubsection{\bfseries{The measurement outcome of $|u_{0}\rangle_{A}$}}}

If the outcome of the measurement is $|u_{0}\rangle_{A}$, the quantum state $|\varphi_{5}\rangle$ is collapsed into
\begin{eqnarray}
    \begin{aligned} 
    |\varphi_{6}\rangle_{Beg}=&(|x_{0}||011\rangle-|x_{0}||110\rangle\\
    &+|x_{1}||000\rangle+|x_{1}||101\rangle)_{Beg}.
\end{aligned}  
\end{eqnarray}

\textbf{Step (\uppercase\expandafter{\romannumeral2})} $Charlie$ who knows the phase information performs a projective measurement on qubit $\textbf{\emph{e}}$ under the orthogonal basis $\{|\nu_{0}\rangle, |\nu_{1}\rangle\}$. Based on the orthogonal basis $\{|\nu_{0}\rangle, |\nu_{1}\rangle\}$, $|\varphi_{5}\rangle$ can be written as
\begin{eqnarray}
    \begin{aligned}   
        |\varphi_{6}\rangle_{Beg}=&e^{-i\theta}|\nu_{0}\rangle_{e}[(x_{1}|0\rangle-x_{0}|1\rangle)_{B}|0\rangle_{g}\\
    &+(x_{0}|0\rangle+x_{1}|1\rangle)_{B}|1\rangle_{g}]\\
    &+|\nu_{1}\rangle_{e}[(x_{1}|0\rangle-x_{0}|1\rangle)_{B}|0\rangle_{g}\\
    &+(x_{0}|0\rangle+x_{1}|1\rangle)_{B}|1\rangle_{g}].
   \end{aligned} 
\end{eqnarray}

\textbf{Step (\uppercase\expandafter{\romannumeral4})}After the measurement of qubit $\textbf{\emph{e}}$, $Charlie$ informs $Bob$ of the result of this measurement. Then $Bob$ measures qubit $\textbf{\emph{g}}$ based on the basis $\{|0\rangle,|1\rangle\}$. Afterthat, the target state $|\psi\rangle=x_{0}|0\rangle+x_{1}|1\rangle$ can be reconstructed by $Bob$ performing the corresponding Puali operation $\sigma_{i}$ on $\textbf{\emph{B}}$ according to the different measurements of two qubits $\textbf{\emph{e}}$ and $\textbf{\emph{g}}$, which is shown in the table below.

\begin{table}[!ht]
    \begin{tabular}{cm{6cm}<{\centering}cm{6cm}<{\centering}cm{6cm}<{\centering}c}
	\hline &$|0\rangle_{g}$&$|1\rangle_{g}$\\
	\hline $|\nu_{0}\rangle_{e}$&$\sigma_{z}$& $I$\\
	\hline $|\nu_{1}\rangle_{e}$&$\sigma_{z}$&$I$\\
	\hline
\end{tabular}
\end{table}

{\centering\subsubsection{\bfseries{The measurement outcome of $|u_{1}\rangle_{A}$}}}

If the outcome of the measurement is $|u_{1}\rangle_{A}$, the quantum state of the system is collapsed into
\begin{eqnarray}
    \begin{aligned}  
    |\varphi_{6}\rangle_{Beg}=&(|x_{1}||011\rangle-|x_{1}||110\rangle\\
    &-x_{0}|000\rangle-|x_{0}||101\rangle)_{Beg}.
\end{aligned}
\end{eqnarray}

\textbf{Step (\uppercase\expandafter{\romannumeral2})} $Charlie$ who knows the phase information performs a phase gate $P(\theta)^{(2)}_{e}$ on qubit $\textbf{\emph{e}}$ and operates a projective measurement on qubit $\textbf{\emph{e}}$ under the orthogonal basis $\{|\nu_{0}\rangle, |\nu_{1}\rangle\}$.
\begin{eqnarray}
    \begin{aligned}   
    |\varphi_{6}\rangle_{Beg}=&|\nu_{0}\rangle_{e}[-(x_{0}|0\rangle+x_{1}|1\rangle)_{B}|0\rangle_{g}\\
    &+(-x_{0}|1\rangle+x_{1}|0\rangle)_{B}|1\rangle_{g}]\\
    &+e^{i\theta}|\nu_{1}\rangle_{e}[-(x_{0}|0\rangle+x_{1}|1\rangle)_{B}|0\rangle_{g}\\
    &+(-x_{0}|1\rangle+x_{1}|0\rangle)_{B}|1\rangle_{g}].
   \end{aligned} 
\end{eqnarray}\\
\\

\textbf{Step (\uppercase\expandafter{\romannumeral4})}After the measurement of qubit $\textbf{\emph{e}}$, $Charlie$ informs $Bob$ of the result of this measurement. Then $Bob$ measures qubit $\textbf{\emph{g}}$ based on the basis $\{|0\rangle,|1\rangle\}$. Afterthat, the target state $|\psi\rangle=x_{0}|0\rangle+x_{1}|1\rangle$ can be reconstructed by $Bob$ performing the corresponding Puali operation $\sigma_{i}$ on $\textbf{\emph{B}}$ according to the different measurements of two qubits $\textbf{\emph{e}}$ and $\textbf{\emph{g}}$, which is displayed in the following chart. And the quantum circuit diagram of posterior measurements corresponding to the measurement result $|1\rangle_{f}$ is displayed in Fig. \ref{Fig.4}.
\begin{table}[!ht]
    \begin{tabular}{cm{6cm}<{\centering}cm{6cm}<{\centering}cm{6cm}<{\centering}c}
	\hline &$|0\rangle_{g}$&$|1\rangle_{g}$\\
	\hline $|\nu_{0}\rangle_{e}$&$I$& $i\sigma_{y}$\\
	\hline $|\nu_{1}\rangle_{e}$&$I$&$i\sigma_{y}$\\
	\hline
\end{tabular}
\end{table}\\

\textbf{Acknowledgements}
This work was supported by National Natural Science Foundation of China (NSFC) under Grant No. 11574041 and No. 12274053.

\textbf{Conflict of Interest}
The authors declare no conflict of interest.

\textbf{Data Availability Statement}
The data that support the findings of this study are available from the corresponding author upon reasonable request.

\bibliographystyle{MSP}
\bibliography{literature}

\end{document}